\documentclass[11pt]{article}
\usepackage{epsfig} 
\newcommand{\kslash}{k \! \! \! /}

\newcommand{\half}{\mbox{\small{$\frac{1}{2}$}}} 
 
\newcommand{\Nf}{N_{\!f}} 
\newcommand{\MSbar}{\overline{\mbox{MS}}} 

\begin{document}
\title{Anomalous dimension of non-singlet quark currents at $O(1/\Nf^2)$ in
QCD} 
\author{D.B. Ali \& J.A. Gracey, \\ Theoretical Physics Division, \\ Department 
of Mathematical Sciences, \\ University of Liverpool, \\ Peach Street, \\ 
Liverpool, \\ L69 7ZF, \\ United Kingdom.} 
\date{} 
\maketitle 
\vspace{5cm} 
\noindent 
{\bf Abstract.} We compute the $O(1/\Nf^2)$ corrections to the flavour
non-singlet quark bilinear currents in QCD in arbitrary spacetime dimensions.
Hence, the anomalous dimension of the QED current $\bar{\psi} \sigma^{\mu\nu} 
\psi$ is deduced at four loops in the $\MSbar$ scheme up to one unknown 
parameter.  

\vspace{-17cm} 
\hspace{13.5cm} 
{\bf LTH 502} 

\newpage 

Recently, the large $\Nf$ method of computing the renormalization group 
functions of a quantum field theory has been extended to the determination of
information in quantum chromodynamics, (QCD), at a {\em new} order of 
expansion, $O(1/\Nf^2)$, \cite{1}. In \cite{1,2} the critical exponents 
corresponding to the anomalous dimensions of the quark field and the quark mass
were computed in $d$-dimensions by exploiting the structure of the field theory
at the $d$-dimensional infrared stable fixed point of the $\beta$-function as 
well as using the relation between QCD and the reduced model known as the 
non-abelian Thirring model, (NATM), \cite{3}. It was demonstrated in \cite{1,2}
that the gluon field strength operator became irrelevant at the fixed point 
allowing one to perform calculations with only the quark gluon vertex and the 
ghost vertex. This important observation, \cite{3}, paved the way for the new 
$O(1/\Nf^2)$ calculations in QCD. Whilst the quark anomalous dimension is a 
gauge dependent quantity, its evaluation in $d$-dimensions in the Landau gauge 
is necessary for other $O(1/\Nf^2)$ computations, since like in explicit 
perturbative calculations the wave function renormalization needs to be 
performed first. However, in additionally providing the $d$-dimensional value 
of the quark mass anomalous dimension, which is a gauge independent quantity, 
it was possible not only to verify the correctness of the recent four loop 
$\MSbar$ perturbative quark mass dimension, \cite{4,5}, at $O(1/\Nf^2)$ but 
also to determine several of the coefficients which will appear in the {\em 
five} loop $\MSbar$ anomalous dimension. Indeed given the huge complexity of 
performing the very high order computations in QCD, insight into the large 
order structure of the renormalization group functions can currently, we 
believe, only be gained by approaches such as the large $\Nf$ expansion coupled
with other techniques. For instance, a recent study by Broadhurst, \cite{6}, of
QED in the quenched approximation involving the Schwinger Dyson equation has 
allowed several new coefficients to be determined in the three and four loop 
$\MSbar$ anomalous dimensions of various operators. By quenched we mean that 
part of the renormalization group functions which does not depend on $\Nf$ and 
hence arises from Feynman diagrams with no electron (or quark in the case of 
QCD) bubbles. Clearly given the amount of resources required for large loop
calculations it would seem that the construction of the high order terms could 
be achieved by the marriage of the large $\Nf$ approach which performs the 
electron or quark loop bubble sums with a method which focuses on the harder 
$\Nf$ $=$ $0$ graphs which have no subdiagrams. Therefore, it is the purpose of
this article to extend the $O(1/\Nf)$ calculation of \cite{7} of the anomalous 
dimension of the (flavour non-singlet) quark currents $\bar{\psi} {\cal A} 
\psi$, where ${\cal A}$ represents a set of $\gamma$-matrices, to $O(1/\Nf^2)$.
Since the case ${\cal A}$ $=$ $1$ corresponds to the quark mass operator, our 
calculation builds on the $O(1/\Nf^2)$ $d$-dimensional computations of 
\cite{1,2}. As a consequence of the result we present here, it will be 
possible, for instance, to deduce three of the four $\MSbar$ coefficients in 
the polynomial in $\Nf$ of the anomalous dimension of the QED operator 
$\bar{\psi} \sigma^{\mu\nu} \psi$, where $\sigma^{\mu\nu}$ $=$ $\half [ 
\gamma^\mu, \gamma^\nu ]$, with the quenched result of \cite{6}. Moreover, 
since the three loop $\MSbar$ anomalous dimension of the general set of 
operators $\bar{\psi} {\cal A} \psi$ has been provided in explicit perturbation
theory in QCD in \cite{8}, there is a non-trivial perturbative check on our 
final result. A final motivation for considering the renormalization of these 
operators rests in the fact that they underly the running of various quantities
in heavy quark effective field theory, \cite{7}, and it is therefore important
to have information on their anomalous dimensions. 
 
We will determine the anomalous dimension of the set of operators
\begin{equation} 
{\cal O}_{(n)} ~=~ \bar{\psi} \Gamma_{(n)}^{\mu_1 \ldots \mu_n} \psi 
\end{equation} 
where $\Gamma^{\mu_1 \ldots \mu_n}_{(n)}$ is defined by 
\begin{equation} 
\Gamma_{(n)}^{\mu_1 \ldots \mu_n} ~=~ \gamma^{[\mu_1} \ldots \gamma^{\mu_n]}
\end{equation} 
and is totally antisymmetric in the Lorentz indices for $n$ $\geq$ $1$. This
particular combination of $\gamma$-matrices is chosen since there will be no
mixing under renormalization and also because the set $\{\Gamma_{(n)}\}$ 
corresponds to a complete $\gamma$-matrix basis in $d$-dimensions, \cite{9}. 
The properties of the matrices $\Gamma_{(n)}$ are well established and given in
\cite{10,11,12}. To compute the large $\Nf$ contribution to the operator 
anomalous dimension one follows the standard critical point approach of 
\cite{13} 
adopted for QCD, \cite{14,2}. This involves using the NATM where there is no
triple or quartic gluon self interactions and the quark and gluon propagators 
are replaced by their leading $d$-dimensional critical point forms,  
\begin{equation}
\psi(k) ~=~ \frac{A\kslash}{(k^2)^{\mu-\alpha}} ~~~,~~~
A_{\mu\nu}(k) ~=~ \frac{B}{(k^2)^{\mu-\beta}}\left[ \eta_{\mu\nu}
- (1-b)\frac{k_\mu k_\nu}{k^2} \right] 
\label{critprops}
\end{equation} 
where $A$ and $B$ are the momentum independent amplitudes, $\alpha$ and $\beta$
are the respective dimensions of the quark and gluon fields and $b$ is the 
covariant gauge fixing parameter. The field dimensions are related to their
anomalous dimensions via 
\begin{equation} 
\alpha ~=~ \mu ~-~ 1 ~+~ \half \eta ~~~,~~~ \beta ~=~ 1 ~-~ \eta ~-~ \chi
\end{equation}
where $\eta$ is the quark anomalous dimension critical exponent, $\chi$ is the
anomalous dimension of the quark gluon vertex and the spacetime dimension $d$
is related to $\mu$ by $d$ $=$ $2\mu$, \cite{14,2}. For the Feynman diagrams 
which arise at the order we consider none involve ghost fields which is why we 
have omitted its critical propagator. Ghost contributions are present 
indirectly in the final critical exponent for the operators ${\cal O}_{(n)}$ 
through the contribution from $\eta$ and the combination $z$ $=$ $A^2B$ which
arises from each vertex. Then the operators ${\cal O}_{(n)}$ are substituted 
into the two point Green's function $\langle \psi(p) {\cal O}_{(n)} 
\bar{\psi}(-p) \rangle$ and the scaling behaviour of the appropriate diagrams 
determined using (\ref{critprops}). In particular a regularization with respect
to $\Delta$ is introduced by shifting the anomalous dimension of the vertex, 
$\beta$ $\rightarrow$ $\beta$ $-$ $\Delta$, and the poles with respect to 
$\Delta$ are removed in a standard renormalization by the appropriate 
renormalization constants. If $p$ is the external momentum then the resulting 
finite Green's function will involve a $\ln p^2$ term. For that Feynman diagram
its coefficient contributes to the overall anomalous dimension critical 
exponent of the operator, \cite{15}. By summing all such contributions at each 
order in $1/\Nf$ one arrives at the full anomalous dimension. As we are 
interested in the $O(1/\Nf^2)$ calculation there are only seven Feynman graphs 
to consider which are illustrated in figures 1 and 2 of \cite{16} as well as 
the three loop graphs of figure 2 of \cite{2} where the operator insertion is 
on a fermionic line which joins to an external vertex.  

Although it was argued in \cite{2} that when computing in the critical point
large $\Nf$ approach, the Landau gauge was the only sensible choice of gauge,
we have chosen to calculate the operator dimension in the Feynman gauge. The 
main reason for this is that since the operators ${\cal O}_{(n)}$ are gauge 
independent, their anomalous dimensions are therefore independent of $b$ and 
hence can be determined in any gauge. We therefore set $b$ $=$ $1$ in 
(\ref{critprops}) to reduce the number of terms in the Feynman integrals and
thereby minimise the amount of computation to be performed. Further, we will
require $\eta$ and $z$ at $O(1/\Nf^2)$ in this gauge. Since we are following
the original skeleton Schwinger Dyson approach of \cite{13}, we record their  
respective values by first setting
\begin{equation} 
z ~=~ \frac{z_1}{\Nf} ~+~ \frac{z_2}{\Nf^2} ~+~ O \left( \frac{1}{\Nf^3} 
\right) ~~~,~~~  
\eta ~=~ \frac{\eta_1}{\Nf} ~+~ \frac{\eta_2}{\Nf^2} ~+~ O \left( 
\frac{1}{\Nf^3} \right) ~.  
\label{zdefns} 
\end{equation}  
Then for arbitrary $b$, \cite{14,2},  
\begin{eqnarray} 
\eta^b_1 &=& \frac{[(2\mu-1)(\mu-2) + b\mu]C_2(R)\eta_0}{(2\mu-1)(\mu-2)T(R)}
{}~~~~,~~~~ z_1 ~=~ \frac{\Gamma(\mu+1)\eta_0}{2(2\mu-1)(\mu-2)T(R)} 
\nonumber \\  
\eta^b_2 &=& \left[ \frac{}{}  
-~ \mu(2 \mu^2 + \mu b - 5 \mu + 2) (\mu - 1) \left[ \hat{\Psi}^2(\mu) 
+ \hat{\Phi}(\mu) \right] C_2(G) \right. \nonumber \\ 
&& \left. ~~+~ \frac{(8 \mu^5 - 92 \mu^4 + 270 \mu^3 - 301 \mu^2 + 124 \mu 
- 12)(2 \mu^2 + \mu b - 5 \mu + 2) \hat{\Psi}(\mu) C_2(G) }{2 (2 \mu - 1) 
(2 \mu - 3) (\mu - 2)} \right. \nonumber \\ 
&& \left. ~~+~ 3\mu(\mu-1)[ \mu b C_2(G) + 2 (2 \mu^2 + \mu b - 5 \mu + 2) 
C_2(R) ] \hat{\Theta}(\mu) \right. \nonumber \\ 
&& \left. ~~-~ [(32 \mu^7 b - 96 \mu^7 + 8 \mu^6 b^2 - 224 \mu^6 b + 912 \mu^6 
- 4 \mu^5 b^3 - 84 \mu^5 b^2 + 704 \mu^5 b - 3360 \mu^5 \right. \nonumber \\
&& \left. ~~~~~~~+ 16 \mu^4 b^3 + 278 \mu^4 b^2 - 1124 \mu^4 b + 6240 \mu^4 
- 19 \mu^3 b^3 - 387 \mu^3 b^2 + 846 \mu^3 b - 6292 \mu^3 \right. \nonumber \\
&& \left. ~~~~~~~+ 6 \mu^2 b^3 + 230 \mu^2 b^2 - 222 \mu^2 b + 3416 \mu^2 - 
48 \mu b^2 - 4 \mu b - 908 \mu + 88) C_2(G) \mu \right. \nonumber \\
&& \left. ~~~~~~~- 8 (4 \mu^5 + 4 \mu^4 b - 32 \mu^4 - 13 
\mu^3 b + 75 \mu^3 + 8 \mu^2 b - 70 \mu^2 - 2 \mu b \right. \nonumber \\
&& \left. ~~~~~~~~~~~~~+ 32 \mu - 6) (2 \mu - 1) (2 \mu - 3) (\mu - 2) C_2(R) ] 
\right. \nonumber \\
&& \left. ~~~~~~~/[4 (2 \mu - 1) (2 \mu - 3) (\mu - 1) (\mu - 2) \mu] \frac{}{}
\right] \frac{C_2(R) \eta_0^2}{2(2\mu-1)^2(\mu-2)^2 T^2(R)} 
\end{eqnarray} 
where the second result derives from the $O(1/\Nf^2)$ solution of the Schwinger
Dyson $2$-point equations for non-zero $b$, and 
\begin{eqnarray} 
z_2 &=& \frac{3\mu(\mu-1)\Gamma(\mu+1) C_2(R) \eta_0^2}{2(2\mu-1)^2(\mu-2)^2 
T^2(R)} \left[ \hat{\Theta}(\mu) ~-~ \frac{1}{(\mu-1)^2} \right] \nonumber \\ 
&& +~ \frac{\mu^2(\mu-1)\Gamma(\mu) C_2(G) \eta_0^2}{4(2\mu-1)^2(\mu-2)^2 
T^2(R)} \left[ 3\hat{\Theta}(\mu) ~-~ \hat{\Psi}^2(\mu) ~-~ \hat{\Phi}(\mu) ~+~ 
\frac{b^2}{4(\mu-1)^2} \right. \nonumber \\ 
&& \left. ~~~~~~~~~~~+~ \frac{(8\mu^5-92\mu^4+270\mu^3-301\mu^2+124\mu-12) 
\hat{\Psi}(\mu)}{2\mu(\mu-1)(2\mu-1)(2\mu-3)(\mu-2)} 
~+~ \frac{(\mu^2+2\mu-2)b}{2(\mu-1)^2\mu^2} \right. \nonumber \\
&& \left. ~~~~~~~~~~~-~ \frac{(16\mu^7-120\mu^6+420\mu^5-776\mu^4+742\mu^3 
-349\mu^2+84\mu-12)}{2(2\mu-1)(2\mu-3)(\mu-1)^2(\mu-2)\mu^2} \right]  
\end{eqnarray} 
where 
\begin{eqnarray}
\eta_0 &=& \frac{(2\mu-1)(\mu-2)\Gamma(2\mu)}{4\Gamma^2(\mu)\Gamma(\mu+1)
\Gamma(2-\mu)} \nonumber \\ 
\hat{\Psi}(\mu) &=& \psi(2\mu-3) ~+~ \psi(3-\mu) ~-~ \psi(1) ~-~ \psi(\mu-1) 
\nonumber\\
\hat{\Phi}(\mu) &=& \psi^\prime(2\mu-3) ~-~ \psi^\prime(3-\mu) ~-~ 
\psi^\prime(\mu-1) ~+~ \psi^\prime(1) \\
\hat{\Theta}(\mu) &=& \psi^\prime(\mu-1) ~-~ \psi^\prime(1) \nonumber
\end{eqnarray}
and $\psi(x)$ is the derivative of the logarithm of the Euler 
$\Gamma$-function.

Since the evaluation of the $O(1/\Nf^2)$ Feynman diagrams follows the standard 
methods of \cite{2,11,14}, we quote the final value of the critical exponent 
for the operator ${\cal O}_{(n)}$ for arbitrary $n$. We find  
\begin{eqnarray} 
\eta^{(n)}_2 &=& \frac{\mu(\mu-1)(2\mu-n-1)(n-1)C_2(R)\eta_0^2} 
{(2\mu-1)^2(\mu-2)^2 T^2(R)} \nonumber \\ 
&& \left[ 2C_2(R) \left( 3\hat{\Theta}(\mu)  
+ \frac{(4\mu^3 - 6\mu^2n - 13\mu^2 + 3\mu n^2 + 2\mu n + 9\mu - n^2 - 3)} 
{\mu^2(\mu-1)^2} \right) \right. \nonumber \\  
&& ~+ \left. C_2(G) \left( \frac{(8\mu^5-92\mu^4+270\mu^3-301\mu^2+124\mu-12) 
\hat{\Psi}(\mu)}{2\mu(\mu-1)(2\mu-1)(2\mu-3)(\mu-2)} ~-~ \hat{\Psi}^2(\mu) ~-~ 
\hat{\Phi}(\mu) \right. \right. \nonumber \\ 
&& ~~~~~~~~~~~~~~~~- \left. \left. [16\mu^6-32\mu^5n-128\mu^5+16\mu^4n^2 
+128\mu^4n+480\mu^4-64\mu^3n^2 \right. \right. \nonumber \\ 
&& \left. \left. ~~~~~~~~~~~~~~~~~~~~- 152\mu^3n-900\mu^3+76\mu^2n^2+48\mu^2n 
+831\mu^2 \right. \right. \nonumber \\ 
&& \left. \left. ~~~~~~~~~~~~~~~~~~~~-24\mu n^2-344\mu+44)] 
/[4\mu(2\mu-1)(2\mu-3)(\mu-1)^2(\mu-2)] \frac{}{} \right) \right] 
\label{op2} 
\end{eqnarray} 
in the notation of (\ref{zdefns}). The value for $\eta^{(n)}_1$ was given in 
\cite{7} and we have correctly reproduced it in our leading order calculation. 
 
There are various checks on the result, (\ref{op2}). First, since the operator
$\bar{\psi} \gamma^\mu \psi$ is a conserved physical current, its anomalous
dimension must vanish at all orders in perturbation theory. The overall factor
of $(n-1)$ which naturally emerges in (\ref{op2}) ensures this. Second, the
result for the quark mass anomalous dimension at $O(1/\Nf^2)$, \cite{1}, is
recovered when $n$ $=$ $0$. The remaining checks are a consequence of comparing
with explicit perturbative calculations since the three loop $\MSbar$ anomalous
dimension for ${\cal O}_{(n)}$ has recently been provided in \cite{8} and is  
\begin{eqnarray} 
\gamma_{(n)}(a) &=& -~ (n - 1)(n - 3)C_2(R) a \nonumber \\ 
&& +~ \left[ 4(n - 15)T(R)\Nf + (18n^3 - 126n^2 + 163n + 291) C_2(G) \right.  
\nonumber \\ 
&& \left. ~~~~~-~ 9(n - 3)(5n^2 - 20n + 1) C_2(R) \right] 
\frac{(n-1)C_2(R)a^2}{18} \nonumber \\ 
&& +~ \left[ \left( 144n^5 - 1584n^4 + 6810n^3 - 15846n^2 + 15933n + 11413 
\right. \right. \nonumber \\ 
&& \left. \left. ~~~~~-~ 216n(n - 3)(n - 4)(2n^2 - 8n + 13)\zeta(3) \right) 
C^2_2(G) \right. \nonumber \\
&& \left. ~~~~~+~ \left( 432n(n - 3)(n - 4)(3n^2 - 12n + 19)\zeta(3) \right.
\right.  \nonumber \\
&& \left. \left. ~~~~~~~~~~~-~ 3(72n^5 - 792n^4 + 3809n^3 - 11279n^2 + 15337n 
+ 1161) \right) C_2(G) C_2(R) \right. \nonumber \\ 
&& \left. ~~~~~+~ \left( 1728(n - 3)\zeta(3) + 8( 3n^3 + 51n^2 - 226n - 278 
) \right) C_2(G) T(R) \Nf \right. \nonumber \\ 
&& \left. ~~~~~-~ \left( 864n(n - 3)(n - 4)(n^2 - 4n + 6)\zeta(3) \right. 
\right. \nonumber \\
&& \left. \left. ~~~~~~~~~~~+~ 18(n - 3)(17n^4 - 136n^3 + 281n^2 - 36n + 129) 
\right) C^2_2(R) \right. \nonumber \\
&& \left. ~~~~~-~ \left( 1728(n - 3) \zeta(3) + 12(17n^3 + n^2 - 326n 
+ 414) \right) C_2(R) T(R) \Nf \right.  \nonumber \\ 
&& \left. ~~~~~+~ 16(13n - 35) T^2(R) \Nf^2 \frac{}{} \right] \frac{(n-1) 
C_2(R) a^3}{108} ~+~ O(a^4) 
\label{pertres} 
\end{eqnarray} 
where $\zeta(z)$ is the Riemann zeta function and $a$ is related to the strong 
coupling constant, $\alpha_s$, by $a$ $=$ $\alpha_s/(4\pi)$. The lower order 
coefficients for $n$ $=$ $2$, for example, had been established in \cite{7,17}
whilst the quenched term at three loop in QED had been given in \cite{6}. To 
compare (\ref{op2}) with (\ref{pertres}) we evaluate the latter at the critical
coupling, $a_c$, in $d$-dimensions and expand in powers of $\epsilon$ and 
$1/\Nf$, where $d$ $=$ $4$ $-$ $2\epsilon$, with \cite{14} 
\begin{eqnarray}
a_c &=& \frac{3\epsilon}{T(R) \Nf} ~+~ \frac{1}{T^2(R) \Nf^2} \left[ 
\frac{33}{4} C_2(G)\epsilon ~-~ 
\left( \frac{27}{4}C_2(R)+\frac{45}{4}C_2(G) \right) \epsilon^2 \right. 
\nonumber \\ 
&& \left. +~ \left( \frac{99}{16}C_2(R) + \frac{237}{32}C_2(G) \right) 
\epsilon^3 ~+~ \left( \frac{77}{16}C_2(R) + \frac{53}{32}C_2(G) \right) 
\epsilon^4 \right. \nonumber \\ 
&& \left. -~ \frac{3\epsilon^5}{256} \left[ (288\zeta(3) + 214) C_2(R) \,+\, 
(480\zeta(3) - 229) C_2(G) \right] \,+\, O(\epsilon^6) \right] \,+\, O \! 
\left( \! \frac{1}{\Nf^3}\! \right) .  
\label{critcoup} 
\end{eqnarray}
To $O(\epsilon^3)$ we find exact agreement between both expansions of the 
critical exponent. Moreover, since (\ref{pertres}) was computed in an arbitrary
covariant gauge and (\ref{op2}) is in agreement with it, we have justified the
choice of Feynman gauge in our calculations.  

Having established the correctness of (\ref{op2}) with all regions of overlap,
we can now determine new information on the higher order terms of 
(\ref{pertres}). Writing the anomalous dimension in terms of its $O(1/\Nf^2)$ 
part as
\begin{eqnarray} 
\gamma_{(n)}(a) &=& -~ (n - 1)(n - 3)C_2(R) a \nonumber \\ 
&& +~ \left[ 4(n - 15)T(R)\Nf + (18n^3 - 126n^2 + 163n + 291) C_2(G) \right.  
\nonumber \\ 
&& \left. ~~~~~-~ 9(n - 3)(5n^2 - 20n + 1) C_2(R) \right] 
\frac{(n-1)C_2(R)a^2}{18} \nonumber \\ 
&& +~ \sum_{r=3}^\infty \left( c_{r0} (T(R))^{r-1} \Nf^{r-1} ~+~ c_{r1} 
(T(R))^{r-2} \Nf^{r-2} C_2(R) \right. \nonumber \\  
&& \left. ~~~~~~~~~~+~ c_{r2} (T(R))^{r-2} \Nf^{r-2} C_2(G) \right) C_2(R) 
a^r ~+~ O \left( \frac{1}{\Nf^3} \right) 
\end{eqnarray} 
and expanding (\ref{op2}) to $O(\epsilon^5)$ using (\ref{critcoup}), we find 
the new $\MSbar$ coefficients  
\begin{eqnarray} 
c_{40} &=& 8(n-1) [ 45 n - 83 - 48 \zeta(3) (n-3) ]/81 \nonumber \\ 
c_{41} &=& 4(n-1) [ 143 n^3 - 1205 n^2 + 2292 n + 228 + 72 (11 n-45) \zeta(3) 
- 648 (n-3) \zeta(4) ]/81 \nonumber \\ 
c_{42} &=& -~ 2(n-1) [130 n^3 - 958 n^2 + 1683 n - 671 + 144 (11 n-45) \zeta(3) 
- 1296 (n-3) \zeta(4) ]/81 \nonumber \\ 
c_{50} &=& 16 (n-1) [ 16(n-15)\zeta(3) - 144(n-3)\zeta(4) + 5(25n-39) ]/243 
\nonumber \\ 
c_{51} &=& 2(n-1) [ - 144 (60 n^3-420 n^2+731 n-297) \zeta(3) 
+ 2592 (11 n-45) \zeta(4) \nonumber \\ 
&& ~~~~~~~~~\, - 20736 (n-3) \zeta(5) 
+ ( 10308 n^3 - 58428 n^2 + 63695 n + 40347) ]/729 \nonumber \\ 
c_{52} &=& (n-1) [ 96 (72 n^3-504 n^2+17 n+2013) \zeta(3) 
- 3456 (11 n-51) \zeta(4) \nonumber \\ 
&& ~~~~~~~~~ + 110592 (n-3) \zeta(5) 
- ( 7128 n^3 - 37416 n^2 + 64723 n - 56001) ]/729 ~.  
\label{newcoeffs} 
\end{eqnarray} 
As part of our motivation was to make use of other techniques aside from 
explicit perturbation theory to deduce higher order terms of the 
renormalization group functions, we can now examine the operator 
${\cal O}_{(2)}$ in QED at four loops. The anomalous dimension in the quenched
approximation is available from \cite{6} and restricting (\ref{newcoeffs}) to
QED, we find that  
\begin{eqnarray} 
\gamma_{(2)}^{\mbox{\footnotesize{QED}}}(a) &=& a ~-~ [ 171 ~+~ 52 \Nf ] 
\frac{a^2}{18} \nonumber \\ 
&& +~ \left[ 3285 ~-~ 3456 \zeta(3) ~+~ 864\zeta(3) \Nf ~+~ 588 \Nf ~-~ 
72 \Nf^2 \right] \frac{a^3}{54} \nonumber \\ 
&& +~ \left[ \frac{3200}{3} \zeta(5) ~-~ \frac{2000}{3}\zeta(3) ~-~ 
\frac{10489}{24} ~+~ c_{43} \Nf \right. \nonumber \\ 
&& \left. ~~~~+\, \left( \frac{4544}{81} \,-\, \frac{736}{9} \zeta(3) \,+\,  
32 \zeta(4) \right) \Nf^2 \,+\, \left( \frac{56}{81} \,+\, 
\frac{128}{27} \zeta(3) \right) \Nf^3 \right] a^4 \,+\, O(a^5) \nonumber \\  
\end{eqnarray}  
where only one unknown coefficient, $c_{43}$, remains to be determined. At 
present there is no next to quenched approximation technique which would allow 
for it to be deduced aside, of course, from an explicit calculation at four 
loops where one would isolate only those Feynman graphs which were linear in 
$\Nf$.  

Finally, we address the issue of the relation of the anomalous dimensions of 
${\cal O}_{(n)}$ to their {\em four} dimensional counterparts. Due to the 
arbitrary dimensional nature of the large $\Nf$ method, we were forced to use
the infinite dimensional $\Gamma_{(n)}$ basis which when restricted to four
dimensions is related to the conventional $\gamma$-matrices. However, one has
to be careful in treating $\gamma^5$ which is a strictly four dimensional 
object and never arises in the generalized basis. For example, setting $n$ $=$ 
$0$, $1$ and $2$ in (\ref{op2}) we recover the correct perturbative structure 
of the respective flavour non-singlet currents. For $n$ $=$ $3$ and $4$ the 
anomalous dimensions $\eta^{(3)}$ and $\eta^{(4)}$ do not correspond to those 
of the axial vector and pseudoscalar currents, \cite{18}. The reason for this 
is that the operator with $\Gamma_{(4)}$, for example, does not exactly project
onto the operator with $\gamma^5$ in four dimensions and in some sense 
incorrectly retains an anticommuting $\gamma^5$ in $d$-dimensions. In the large
$\Nf$ approach the method to properly account for the presence of $\gamma^5$ 
has been developed in \cite{19} based on earlier observations of \cite{20}. 
Like perturbation theory, \cite{18}, one has to add an additional critical 
exponent which would correspond to a finite renormalization constant in the  
perturbative approach. However, the result of \cite{19} only concerned 
$O(1/\Nf)$ calculations and therefore we extend the method here to the next 
order since it will also be important in future large $\Nf$ calculations of 
other operators which contain a $\gamma^5$ such as four quark operators. As in 
\cite{19} we define a critical exponent corresponding to the finite 
renormalization by
\begin{equation} 
\eta^{\mbox{\footnotesize{fin}}} ~=~ \eta^{(4-n)} ~-~ \eta^{(n)} ~.  
\end{equation}  
This choice is motivated by \cite{18,8} in that since the currents are 
non-singlet, and hence anomaly free, this condition will restore the 
four dimensional properties of $\gamma^5$ in the renormalization of the 
operators. Hence, from (\ref{op2}) we find,
\begin{eqnarray}  
\eta^{\mbox{\footnotesize{fin}}} &=& -~ \frac{8(n-2)C_2(R)\eta_0} 
{(2\mu-1)T(R)\Nf} \nonumber \\  
&& -~ \left[ \frac{}{} 24 \mu (\mu - 1) \hat{\Theta}(\mu) C_2(R) ~-~  
4 \mu (\mu - 1) \left( \hat{\Psi}^2(\mu) + \hat{\Phi}(\mu) \right) C_2(G)  
\right. \nonumber \\ 
&& \left. ~~~~~+~ \frac{2 (8 \mu^5 - 92 \mu^4 + 270 \mu^3 - 301 \mu^2 + 124 \mu 
- 12) \hat{\Psi}(\mu) C_2(G)}{(2 \mu - 1) (2 \mu - 3) (\mu - 2)} \right.
\nonumber \\
&& \left. ~~~~~-~ [16 \mu^6 - 224 \mu^5 + 32 \mu^4 n^2 - 128 \mu^4 n 
+ 1104 \mu^4 - 128 \mu^3 n^2 + 512 \mu^3 n \right. \nonumber \\ 
&& \left. ~~~~~~~~~- 2316 \mu^3 + 152 \mu^2 n^2 - 608 \mu^2 n + 2115 \mu^2 
- 48 \mu n^2 + 192 \mu n \right. \nonumber \\ 
&& \left. ~~~~~~~~~- 704 \mu + 44] C_2(G)/[(2 \mu - 1)(2 \mu - 3)(\mu - 1) 
(\mu - 2)] \right. \nonumber \\ 
&& \left. ~~~~~+~ 8 [4 \mu^3 - 31 \mu^2 + 6 \mu n^2 - 24 \mu n + 60 \mu - 2 n^2
\right. \nonumber \\
&& \left. ~~~~~~~~~~~+ 8 n - 18] C_2(R)/[\mu(\mu - 1)] \frac{}{} \right] 
\frac{(n-2)C_2(R) \eta_0^2}{(2\mu-1)^2 (\mu-2) T^2(R) \Nf^2} ~+\, O \! \left( 
\! \frac{1}{\Nf^3} \right)  
\label{findef} 
\end{eqnarray} 
which is only relevant when $n$ $=$ $3$ or $4$. Adding (\ref{findef}) to 
(\ref{op2}) for these values one correctly recovers the critical exponents for 
the axial vector and pseudoscalar currents which agree with the known four
dimensional perturbative results, \cite{18}. 

We conclude with various remarks. First, we have demonstrated that it is 
possible to marry results from several techniques aside from perturbation 
theory to determine the structure of a renormalization group function at four 
loops in $\MSbar$, up to one unknown parameter. Second, we focused on the
flavour non-singlet currents. For even $n$ the results are also valid for the
flavour singlet currents since the extra graphs which arise when the operator
${\cal O}_{(n)}$ is inserted in a closed quark loop trivially vanish as they 
involve a trace over an odd number of $\gamma$-matrices. The study of the 
flavour singlet currents at $O(1/\Nf^2)$ for odd $n$ is more involved than the 
non-singlet computation. Due to the nature of the $1/\Nf$ expansion the largest
loop order one would have to consider is {\em five} loops. However, similar 
diagrams arise in the computation of the quark gluon vertex anomalous dimension
at $O(1/\Nf^2)$ and since that more fundamental calculation has not yet been 
performed, the formalism and {\em integration} techniques necessary to handle 
the five and lower loop diagrams in the critical point approach using the 
propagators (\ref{critprops}) are not yet available. Therefore, to extend 
(\ref{op2}) to the singlet case, prior to treating the problem of correctly 
accounting for the axial vector anomaly, which is an issue in its own right, 
a substantial amount of additional calculations would be required which is 
beyond the scope of the present article.  

\vspace{1cm} 
\noindent 
{\bf Acknowledgements.} This work was supported in part by a {\sc PPARC}
Research Studentship, (DBA).

\end{document}